# Maximum signal-to-noise ratio enhancement by averaging under a limited measurement time

YingCheng Zhou,[1,2,a] Kosuke Minami,[1] Genki Yoshikawa,[1,2] and Gaku Imamura[1,3,b]

[1] *Research Center for Macromolecules and Biomaterials, National Institute for Materials Science (NIMS), 1-1 Namiki, Tsukuba, Ibaraki 305-0044 Japan*

[2] *Materials Science and Engineering, Graduate School of Pure and Applied Science, University of Tsukuba, 1-1-1 Tennodai, Tsukuba, Ibaraki 305-8571, Japan*

[3] *Graduate School of Information Science and Technology, Osaka University, 1-2 Yamadaoka, Suita, Osaka 565-0871 Japan*

Averaging through repetitive measurement is a ubiquitous strategy for improving signal-to-noise ratio (SNR) and is commonly assumed to yield a $\sqrt{N}$ enhancement with the number of repetitions $N$. This assumption, however, implicitly requires the signal amplitude to be independent of measurement duration. This condition does not generally hold in dynamical sensing systems with finite response time and a fixed measurement time. We derive a closed-form expression for the SNR enhancement factor by analytically accounting for the competition between statistical noise reduction and dynamical signal attenuation, and demonstrate the existence of a strict upper bound on the SNR enhancement. The enhancement factor is a non-monotonic function of $N$ with a well-defined maximum at an optimal repetition number, beyond which further averaging degrades the SNR. Moreover, below a threshold set by the ratio of measurement time to response time, averaging yields no enhancement at all. These two regimes delimit where the conventional $\sqrt{N}$ law breaks down. Experimental validation using nanomechanical gas sensing, with two receptor–analyte systems deliberately chosen to bracket this enhancement transition, confirms the theoretical predictions. Our results show that measurement time is a finite resource to be optimally partitioned between signal accumulation and averaging, and provide a quantitative guideline for selecting the repetition number in time-constrained sensing such as real-time and repetitive gas or odor detection.

Gas sensors have attracted considerable attention[1–4] for applications ranging from environmental monitoring and industrial process control to healthcare diagnostics and food-quality assessment.[5–7] For precise and accurate measurements, it is necessary to detect weak sensing responses buried in noise, which makes improving the signal-to-noise ratio (SNR) central to sensor performance. Averaging over repetitive measurements is a fundamental technique in physical measurement and signal processing, widely employed to recover weak signals from noisy measurements. Under the assumption of statistically independent noise, averaging is commonly expected to improve the SNR in proportion to the square root of the number of repetitions $\sqrt{N}$,[8–10] and is often regarded as universally beneficial. This principle underpins measurement strategies across diverse sensing platforms, from optical trace-gas spectroscopy to chemical gas sensing,[2,11–13] particularly when the signal amplitude approaches or falls below the detection limit. Averaging cannot be extended indefinitely in practice because low-frequency instrumental drift eventually dominates and further averaging becomes counterproductive, as captured by Allan-variance analysis.[14,15] This ceiling, however, is extrinsic to the measurement principle, and below it the $\sqrt{N}$ scaling is treated as an unconditional benefit.

The $\sqrt{N}$ scaling, however, relies on an implicit but restrictive assumption, namely that the signal amplitude associated with each repetition is independent of the measurement duration. In many practical sensing scenarios, this assumption does not hold. Many sensors, including chemical, biological, and nanomechanical sensors,[16,17] respond on a finite timescale set by their intrinsic dynamics.[2,3,18,19] When the total available measurement time is fixed, increasing the number of repetitions necessarily shortens the duration of each measurement cycle, leading to incomplete system response and attenuation of the signal amplitude. As a result, statistical noise reduction through averaging competes directly with dynamical signal loss, as shown in Fig. 1. This trade-off is particularly relevant to repetitive sensing and other signal-enhancement protocols, which are increasingly adopted to improve robustness and reliability,[20,21] yet the number of repetitions is typically chosen empirically, without a principled optimum.

In this work, we investigate the consequences of this competition and establish the maximum SNR enhancement achievable by repetitive measurement under a limited total measurement time, along with the repetition number that attains it. We derive a closed-form expression for the SNR enhancement factor under a fixed total measurement time, explicitly incorporating the finite response time of the sensing system. In contrast to the extrinsic drift ceiling noted above, the limit we identify is intrinsic, arising from the finite dynamical response of the sensor rather than from the noise. We validate these predictions experimentally using nanomechanical gas sensors as a representative dynamical platform, with two receptor–analyte systems that bracket the predicted transition. Our results provide a concrete guideline for when, and how many times, to repeat a measurement.

We consider repetitive measurements performed within a fixed total measurement time $2\tau$, divided into $N$ identical cycles each with a symmetrical active and recovery duration $T = \tau/N$. While increasing $N$ reduces the noise amplitude through statistical averaging by a factor of $\sqrt{N}$, shortening the duration of each cycle leads to incomplete system response and consequently attenuation of the signal amplitude. Because the sensing response develops on a finite timescale, we model it with first-order kinetics.

[a] Electronic mail: ZHOU.Yingcheng@nims.go.jp

[b] Electronic mail: IMAMURA.Gaku@nims.go.jp

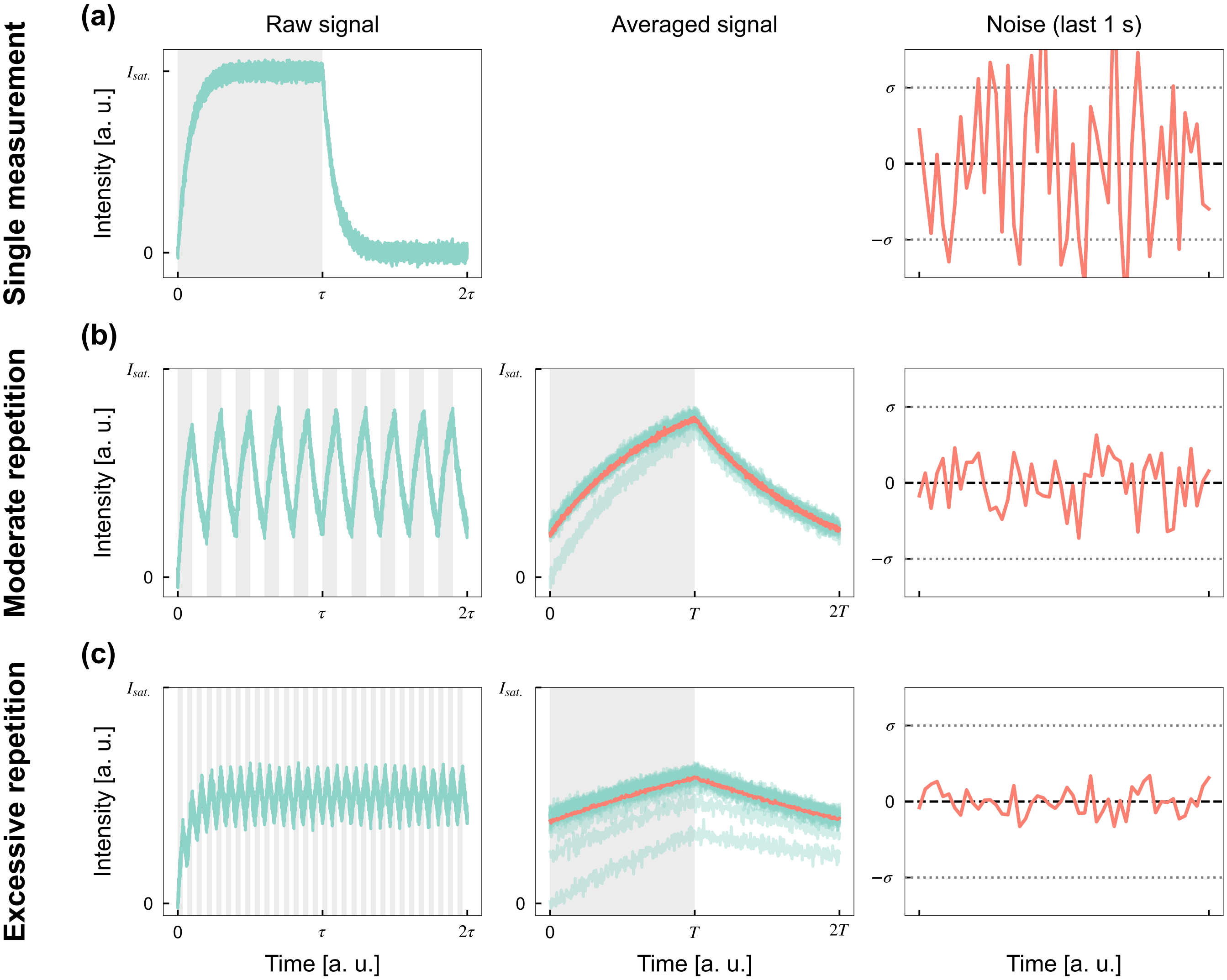

FIG. 1. Schematic illustration of repetitive measurement under a fixed total measurement time. (a) A single long measurement allows the sensor response to fully develop, yielding a large signal amplitude. (b) For a moderate repetition number, noise is reduced by averaging while the signal remains sufficiently strong, resulting in an optimal signal-to-noise ratio. (c) For excessive repetition, each measurement becomes too short for the system to respond, leading to severe signal attenuation and a degradation of signal-to-noise ratio despite further noise reduction. In each row, the three columns show, from left to right, the raw signal over the full measurement time (the injection half of each cycle is shaded), the individual cycles overlaid with their average, and a zoom on the noise in the last part of the averaged cycle. The noise traces are shown after subtracting a linear baseline trend, so that they represent the random noise alone. The dotted lines mark the single-measurement noise level $\pm\sigma$ as a fixed reference across rows.

$$\frac{\mathrm{d}I}{\mathrm{d}t} = \frac{1}{\tau_s}(I_{sat.} - I). \tag{1}$$

where $I$ is the signal intensity, $\tau_s$ is the response time constant of the system, and $I_{sat.}$ is the saturated signal intensity, i.e., the intensity that the response approaches after the system reaches its steady state. In the sorption-based sensors used in our experiments, $\tau_s$ is the sorption time constant.[22,23] Assuming the external stimulus amplitude is independent of time, the analytical solution for the signal response under repetitive measurement can be obtained recursively. Minami *et al.* provide the full derivation.[23] The signal amplitude $\Delta I$ of each measurement cycle is defined as the difference between the signal at the end and at the beginning of the active phase. In the limit $t \longrightarrow \infty$, where the repetitive response settles into a periodic steady state, this per-cycle amplitude is given by

$$\Delta I_{t\to\infty} = I(t_{2n-1}) - I(t_{2(n-1)}) = I_{sat.} \tanh\left(\frac{T}{2\tau_s}\right). \tag{2}$$

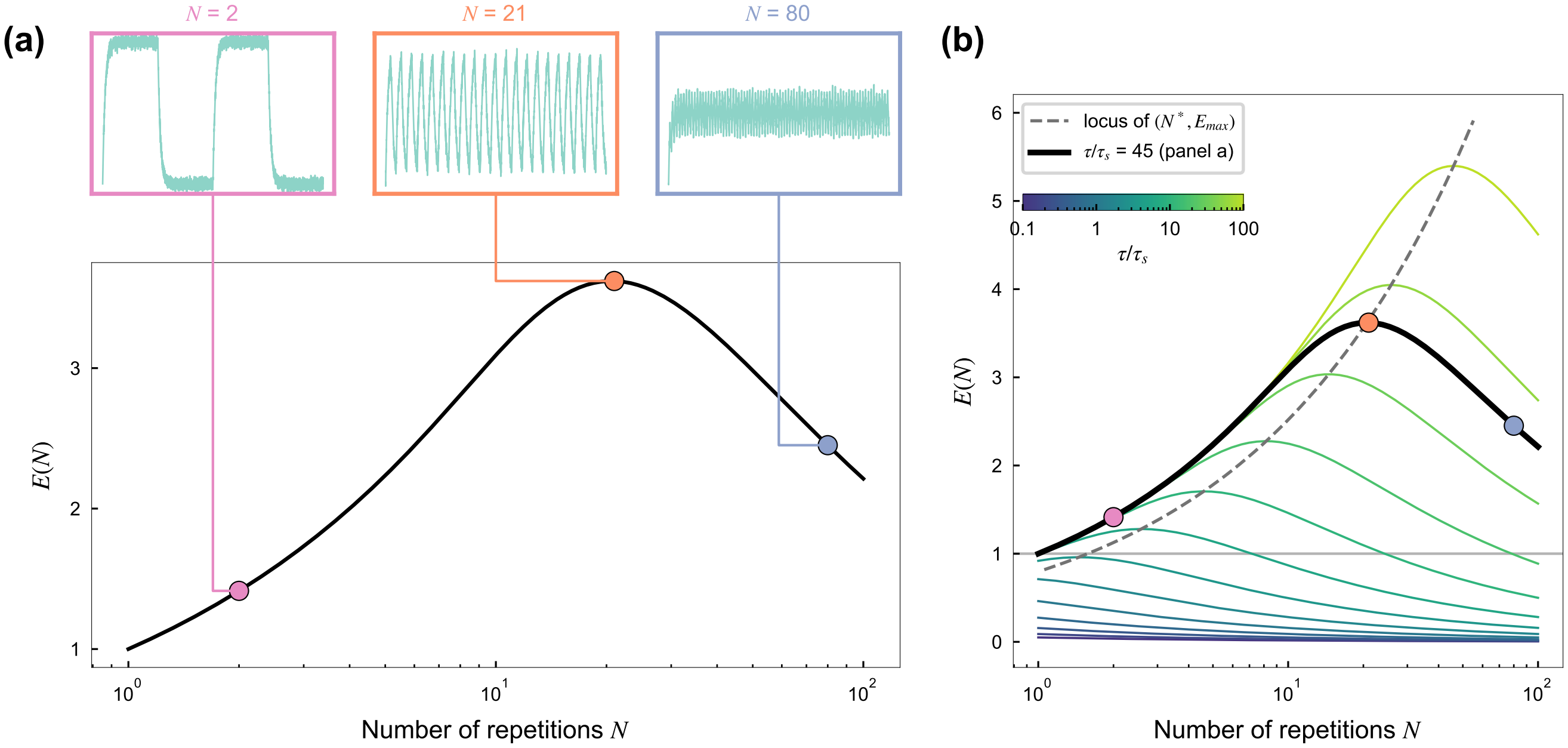

FIG. 2. Numerical analysis of the enhancement factor. (a) Enhancement factor as a function of the number of repetitions. The enhancement factor exhibits a non-monotonic trend with increasing repetition number, demonstrating the existence of an optimal repetition number. Insets show three representative sensing responses at the corresponding repetition numbers, where the signal amplitude decreases monotonically. (b) Theoretical enhancement factor as a function of repetition number for different ratios of total measurement time to sensor response time. Larger ratios yield larger optimal enhancement factors.

The SNR of a single measurement cycle is represented as

$$\mathrm{SNR}_1 = \frac{\Delta I}{\sigma}. \tag{3}$$

With $N$ repetitive measurements and averaging, the SNR becomes

$$\mathrm{SNR}_N = \frac{\Delta I}{\frac{\sigma}{\sqrt{N}}} = \frac{\Delta I\sqrt{N}}{\sigma}. \tag{4}$$

The enhancement factor $E(N)$ is defined as

$$E(N) = \frac{\mathrm{SNR}_N}{\mathrm{SNR}_1} = \sqrt{N}\tanh\left(\frac{\tau}{2\tau_s N}\right). \tag{5}$$

As a result, the enhancement factor depends solely on the response dynamics of the sensing system and is independent of the noise amplitude $\sigma$. $E(N)$ is a non-monotonic function of $N$ and exhibits a well-defined unique maximum at an optimal repetition number, where the gain from noise reduction is exactly balanced by the loss in signal amplitude, as shown in Fig. 2(a). For sufficiently large $N$, the duration of each cycle becomes too short for meaningful signal accumulation, and the loss in signal amplitude outweighs the statistical noise reduction, causing the SNR to decrease with increasing $N$.

In Eq. 5, $\tau_s$ is determined by the receptor–analyte combination and is largely fixed for a given measurement, whereas the total measurement time $\tau$ is set by the experimenter. The behavior of $E(N)$ then reveals two distinct ways in which the conventional $\sqrt{N}$ law breaks down. First, for a fixed $\tau/\tau_s$, $E(N)$ is non-monotonic and attains its maximum at an optimal repetition number $N^*$ [Fig. 2(a)]. Beyond $N^*$, each cycle becomes too short for the response to develop, and the loss in signal amplitude outweighs the statistical noise reduction, so that further averaging degrades the SNR. Second, averaging is beneficial only above a threshold in $\tau/\tau_s$ [Fig. 2(b)]. When the total measurement time is comparable to or shorter than the response time, the system cannot approach steady state within any subdivision of the available time, and $E(N) < 1$ for every $N$. For the single-relaxation model of Eq. 5, this threshold is $\tau/\tau_s \approx 3$. Below it, repetition cannot improve the SNR regardless of $N$. These two regimes together delimit a finite window, in both the repetition number and the time-constant ratio, within which repetitive averaging genuinely enhances the SNR, and outside of which the $\sqrt{N}$ law fails, either by saturation and decline beyond $N^*$ or by yielding no enhancement whatsoever below the threshold.

We experimentally validate these predictions using nanomechanical membrane-type surface stress sensors (MSS)[17,24] as a representative dynamical platform. Details of the measurement setup are given in the supplementary material. Polymer-functionalized MSS were exposed to analyte vapor at a constant maximum concentration, and repetitive measurements were performed under a fixed total measurement time of $2\tau = 360$ s while systematically varying the repetition number $N$ from 1 to 90.[20] To probe both regimes identified above, we deliberately selected two receptor–analyte combinations that bracket the enhancement transition, specifically a fast-sorption case (poly($\varepsilon$-caprolactone) (PCL) receptor exposed to $n$-nonane) and a slow-sorption case (poly(methyl methacry-late) (PMMA) receptor exposed to methanol), the latter having a sorption time constant $\tau_s$ about twice as long. For these polymer receptor layers, the sorption-induced signal is governed by two characteristic time constants (the sorption time $\tau_s$ and a stress relaxation time $\tau_r$), so that the enhancement factor takes a more elaborate form than Eq. 5, depending on both time constants and a modulus ratio.[23] It remains non-monotonic in $N$ and still exhibits a well-defined optimal repetition number, so the qualitative picture established from the single-relaxation model of Eq. 5 carries over. The two time constants do, however, shift the enhancement threshold, which then depends not only on $\tau/\tau_s$ but also on the relaxation dynamics and modulus ratio of the layer. Its derivation and the resulting threshold map are given in the supplementary material. The MSS output voltage $V$ is the experimental realization of the model intensity $I$, so that its per-cycle amplitude $\Delta V$ corresponds to $\Delta I$. For each $N$, the signal amplitude $\Delta V$ and noise level $\sigma$ were extracted from the averaged response,[20] and the parameters obtained from a single measurement were used to predict $E(N)$ throughout the experiment. The results, together with the theoretical predictions, are shown in Fig. 3.

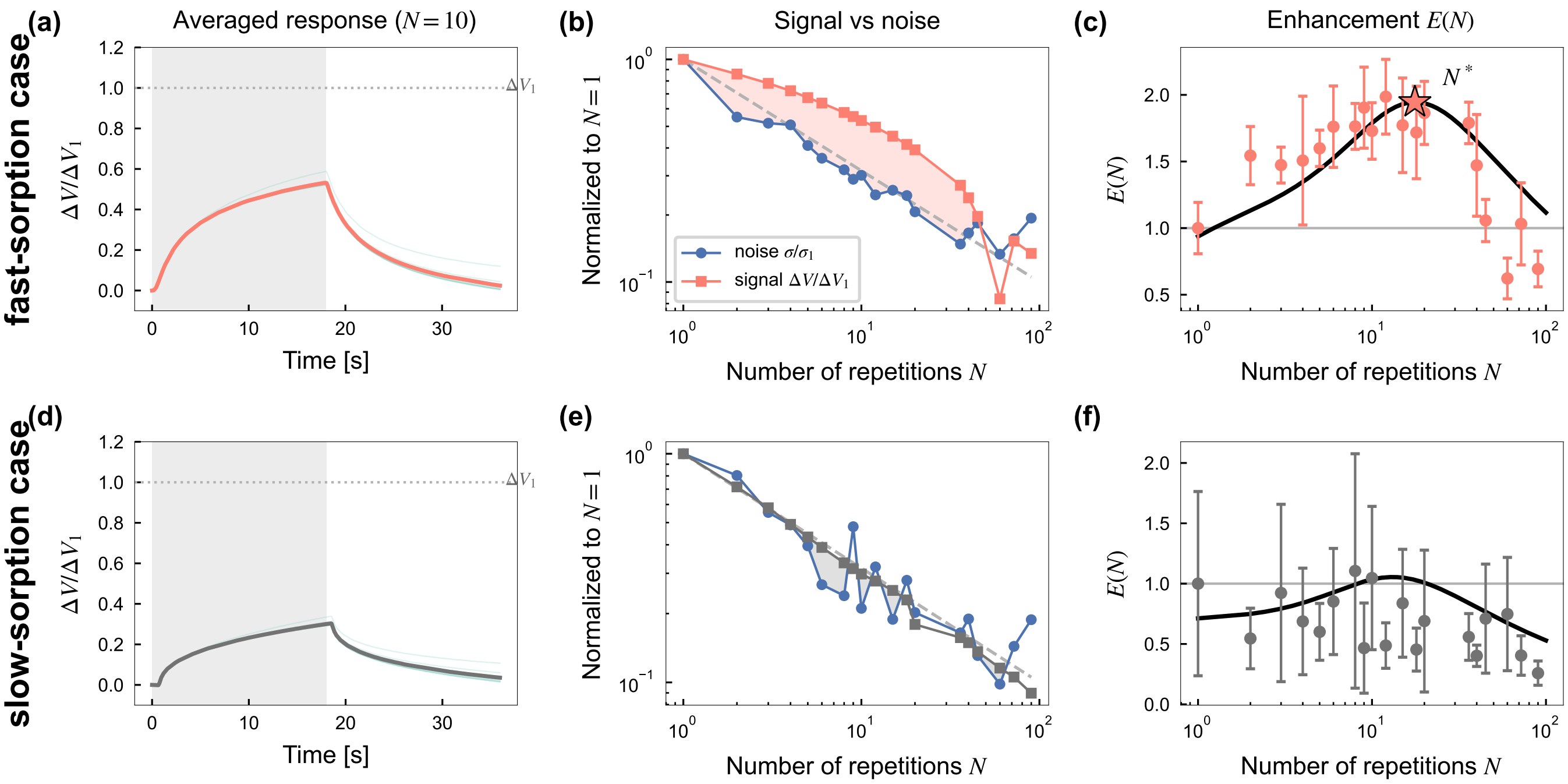

FIG. 3. Experimental validation of the theoretical prediction, comparing a fast-sorption case (top row) and a slow-sorption case (bottom row) measured under the same fixed total measurement time. (a,d) Averaged sensor response for one cycle at $N = 10$, normalized to the single-injection amplitude $\Delta V_1$ (dotted line). The slow-sorption case builds up much less signal per cycle. (b,e) Normalized signal retention $\Delta V/\Delta V_1$ (squares) against the $1/\sqrt{N}$ noise reduction (circles, ideal shown dashed). The shaded gap between them is the net enhancement, which the fast-sorption case opens but the slow-sorption case does not. (c,f) Measured enhancement factor $E(N)$ (points, error bars are standard deviations from five independent measurements) compared with the theoretical prediction (line), computed from the viscoelastic enhancement factor including both the sorption ($\tau_s$) and relaxation ($\tau_r$) time constants. The fast-sorption case rises to a clear maximum at an optimal repetition number $N^*$, whereas within the noise the slow-sorption case shows no meaningful enhancement and hovers around $E = 1$.

As shown in Fig. 3, both the fast-sorption (PCL/$n$-nonane) and slow-sorption (PMMA/methanol) cases exhibit the two competing effects underlying repetitive measurement. The noise level decreases with $N$ following the expected $1/\sqrt{N}$ scaling [Figs. 3(b) and 3(e)], confirming that statistical averaging operates as anticipated, whereas the signal amplitude $\Delta V$ systematically decreases because each cycle becomes too short for the sorption response to develop [Figs. 3(a) and 3(d)]. Whether averaging yields a net gain is therefore set by how fast the signal is lost relative to the $1/\sqrt{N}$ noise reduction.

In the fast-sorption case, the fast-responding receptor retains enough signal per cycle that the noise reduction prevails over a finite range of $N$. The signal-retention curve stays above the $1/\sqrt{N}$ noise floor [Fig. 3(b)], and the measured enhancement factor rises to a clear maximum of $E \approx 1.9$ at $N^* \approx 18$ before declining [Fig. 3(c)], meaning that optimal partitioning of the measurement time nearly doubles the SNR relative to a single measurement. Both the location of $N^*$ and the peak enhancement are in good quantitative agreement with the theoretical prediction, confirming the optimal-repetition regime. Conversely, increasing $N$ well beyond $N^*$ drives $E$ below unity, so that excessive averaging actually degrades the SNR compared with no averaging at all.

In the slow-sorption case, by contrast, the slow-responding receptor accumulates so little signal per cycle that $\Delta V$ falls off essentially as fast as the noise. The signal-retention and noise curves nearly coincide and leave no net gap [Fig. 3(e)]. Within the measurement uncertainty the enhancement factor shows no meaningful improvement and hovers around $E = 1$ for all $N$ [Fig. 3(f)], directly confirming the predicted no-enhancement regime below the threshold. Minor deviations at very large $N$ in both cases are attributed to the finite response time of the gas delivery system (detailed in the supplementary material), which introduces additional signal attenuation not captured by the idealized model. These deviations do not affect the position of $N^*$, which is governed by the intrinsic sensor dynamics.

The upper bound identified here differs fundamentally from the well-known practical ceiling on averaging set by low-frequency instrumental drift.[15] Whereas drift is an extrinsic effect that can in principle be reduced by better instrumentation, the limit we derive is intrinsic to the finite dynamical response of the sensor and independent of the noise amplitude. The optimal repetition number is governed by the system response time and the available measurement window, rather than by noise amplitude, underscoring the dynamical origin of the limitation.

This conclusion is not specific to nanomechanical gas sensors but extends to any sensing system in which the signal is governed by time-dependent sensing dynamics within a finite measurement time, including chemical and biological sensors.[19] In such systems, our closed-form criterion specifies both whether averaging is worthwhile and, if so, the optimal number of repetitions from an independently measured response time, turning the choice of $N$ from an empirical convention into a designed parameter.[20] Adaptive adjustment based on the remaining measurement time may further be used to approach the theoretical SNR limit.

In conclusion, we have identified a fundamental upper bound on SNR enhancement by averaging under a finite measurement duration. The competition between statistical noise reduction and dynamical signal attenuation makes the enhancement non-monotonic in the repetition number, with a well-defined optimum, and precludes any benefit below a response-time threshold. More broadly, our framework treats measurement time as a finite resource to be optimally partitioned between signal accumulation and averaging, providing a quantitative guideline for designing repetitive measurements in any dynamical sensing system constrained by time or sample availability.

See the supplementary material for details of theoretical derivations, experimental setup, and additional data analysis.

This work was supported by JST SPRING, Grant Number JPMJSP2124, and by JSPS KAKENHI Grant Number JP24K08820.

## AUTHOR DECLARATION

### Conflict of Interest

K.M., G.Y., and G.I. are inventors on Japanese Patent No. JP7761314. Y.Z. has no conflicts to disclose.

### Author Contributions

**Yingcheng Zhou:** Conceptualization, Formal analysis, Investigation, Writing - original draft. **Kosuke Minami:** Conceptualization, Writing - review & editing, Supervision. **Genki Yoshikawa:** Writing - review & editing, Supervision. **Gaku Imamura:** Conceptualization, Writing - review & editing, Supervision, Funding acquisition.

## DATA AVAILABILITY

The data that support the findings of this study are available from the corresponding author upon reasonable request.

## REFERENCES

[1] K. Persaud and G. Dodd, "Analysis of discrimination mechanisms in the mammalian olfactory system using a model nose," Nature **299**, 352–355 (1982).

[2] F. Röck, N. Barsan, and U. Weimar, "Electronic Nose: Current Status and Future Trends," Chem. Rev. **108**, 705–725 (2008).

[3] C. Wang, L. Yin, L. Zhang, D. Xiang, and R. Gao, "Metal Oxide Gas Sensors: Sensitivity and Influencing Factors," Sensors **10**, 2088–2106 (2010).

[4] W. Zhang, W. Tang, Z. Wan, and Z. Fan, "Advanced electronic noses for future robotic olfaction," Npj Robot. **4**, 11 (2026).

[5] A. Khan, D. Schaefer, L. Tao, D. J. Miller, K. Sun, M. A. Zondlo, W. A. Harrison, B. Roscoe, and D. J. Lary, "Low Power Greenhouse Gas Sensors for Unmanned Aerial Vehicles," Remote Sens. **4**, 1355–1368 (2012).

[6] Y. Y. Broza, P. Mochalski, V. Ruzsanyi, A. Amann, and H. Haick, "Hybrid Volatolomics and Disease Detection," Angew. Chem. Int. Ed. **54**, 11036–11048 (2015).

[7] Y. Saeki, N. Maki, T. Nemoto, K. Inada, K. Minami, R. Tamura, G. Imamura, Y. Cho-Isoda, S. Kitazawa, H. Kojima, G. Yoshikawa, and Y. Sato, "Lung cancer detection in perioperative patients' exhaled breath with nanomechanical sensor array," Lung Cancer **190**, 107514 (2024).

[8] M. P. Klein and G. W. Barton, "Enhancement of Signal-to-Noise Ratio by Continuous Averaging: Application to Magnetic Resonance," Rev. Sci. Instrum. **34**, 754–759 (1963).

[9] R. R. Ernst and W. A. Anderson, "Application of Fourier Transform Spectroscopy to Magnetic Resonance," Rev. Sci. Instrum. **37**, 93–102 (1966).

[10] P. Horowitz and W. Hill, *The Art of Electronics*, third edition, 21st printing with corrections ed. (Cambridge University Press, Cambridge, New York, 2024).

[11] J. M. Langridge, S. M. Ball, A. J. L. Shillings, and R. L. Jones, "A broadband absorption spectrometer using light emitting diodes for ultrasensitive, *in situ* trace gas detection," Rev. Sci. Instrum. **79**, 123110 (2008).

[12] J. Li, H. Deng, J. Sun, B. Yu, and H. Fischer, "Simultaneous atmospheric CO, $N_2O$ and $H_2O$ detection using a single quantum cascade laser sensor based on dual-spectroscopy techniques," Sens. Actuators B Chem. **231**, 723–732 (2016).

[13] R. Bauer, G. Stewart, W. Johnstone, E. Boyd, and M. Lengden, "3D-printed miniature gas cell for photoacoustic spectroscopy of trace gases," Opt. Lett. **39**, 4796 (2014).

[14] D. Allan, "Statistics of atomic frequency standards," Proc. IEEE **54**, 221–230 (1966).

[15] P. Werle, R. Mücke, and F. Slemr, "The limits of signal averaging in atmospheric trace-gas monitoring by tunable diode-laser absorption spectroscopy (TDLAS)," Appl. Phys. B Photophysics Laser Chem. **57**, 131–139 (1993).

[16] H. P. Lang, M. Hegner, and C. Gerber, "Cantilever array sensors," Mater. Today **8**, 30–36 (2005).

[17] K. Minami, G. Imamura, R. Tamura, K. Shiba, and G. Yoshikawa, "Recent Advances in Nanomechanical Membrane-Type Surface Stress Sensors towards Artificial Olfaction," Biosensors **12**, 762 (2022).

[18] M. Galvani, S. Freddi, and L. Sangaletti, "Disclosing Fast Detection Opportunities with Nanostructured Chemiresistor Gas Sensors Based on Metal Oxides, Carbon, and Transition Metal Dichalcogenides," Sensors **24**, 584 (2024).

[19] B. McCann, B. Tipper, S. Shahbeigi, M. Soleimani, M. Jabbari, and M. Nasr Esfahani, "A Review on Perception of Binding Kinetics in Affinity

Biosensors: Challenges and Opportunities," ACS Omega **10**, 4197–4216 (2025).

[20]G. Imamura, K. Minami, and G. Yoshikawa, "Repetitive Direct Comparison Method for Odor Sensing," Biosensors **13** (2023), 10.3390/bios13030368.

[21]M.-Q. Feng, T. Yildirim, K. Minami, K. Shiba, and G. Yoshikawa, "Sensing signal augmentation by flow rate modulation of carrier gas for accurate differentiation of complex odours," Sci. Technol. Adv. Mater. **25**, 2408212 (2024).

[22]M. J. Wenzel, F. Josse, S. M. Heinrich, E. Yaz, and P. G. Datskos, "Sorption-induced static bending of microcantilevers coated with viscoelastic material," J. Appl. Phys. **103**, 064913 (2008).

[23]K. Minami, K. Shiba, and G. Yoshikawa, "Sorption-induced static mode nanomechanical sensing with viscoelastic receptor layers for multistep injection-purge cycles," J. Appl. Phys. **129**, 124503 (2021).

[24]G. Yoshikawa, T. Akiyama, S. Gautsch, P. Vettiger, and H. Rohrer, "Nanomechanical Membrane-type Surface Stress Sensor," Nano Lett. **11**, 1044–1048 (2011).

**Supplementary Materials**

# Maximum signal-to-noise ratio enhancement by averaging under a limited measurement time

YingCheng Zhou,[1,2,a] Kosuke Minami,[1] Genki Yoshikawa,[1,2] and Gaku Imamura[1,3,b]

[1] *Research Center for Macromolecules and Biomaterials, National Institute for Materials Science (NIMS), 1-1 Namiki, Tsukuba, Ibaraki 305-0044 Japan*

[2] *Materials Science and Engineering, Graduate School of Pure and Applied Science, University of Tsukuba, 1-1-1 Tennodai, Tsukuba, Ibaraki 305-8571, Japan*

[3] *Graduate School of Information Science and Technology, Osaka University, 1-2 Yamadaoka, Suita, Osaka 565-0871 Japan*

[a] Electronic mail: ZHOU.Yingcheng@nims.go.jp

[b] Electronic mail: IMAMURA.Gaku@nims.go.jp

## CONTENTS

## I. DERIVATION OF FIRST-ORDER RESPONSE

We consider a general first-order response system described by the following differential equation

$$\frac{\mathrm{d}I}{\mathrm{d}t} = \frac{1}{\tau_s}\left(I_{sat.} - I\right) \tag{S1}$$

where $I$ is the signal intensity, $I_{sat.}$ is the saturation intensity, and $\tau_s$ is the time constant of the system.

The signal intensity $I(t)$ in the repetitive measurement is given by[1]

$$I(t) = \begin{cases} 0, & t = 0 \quad \text{(S2a)} \\ I_{sat.}\left[1 - e^{-\frac{t-t_0}{\tau_s}} \sum_{i=0}^{2(n-1)} \left(-e^{\frac{T}{\tau_s}}\right)^i\right], & t_{2(n-1)} \leq t < t_{2n-1} \quad \text{(S2b)} \\ I_{sat.}\left[e^{-\frac{t-t_0}{\tau_s}} \sum_{i=0}^{2n-1} \left(-e^{\frac{T}{\tau_s}}\right)^i\right]. & t_{2n-1} \leq t < t_{2n} \quad \text{(S2c)} \end{cases}$$

Here, we define the signal amplitude ($\Delta I$) as the difference between the initial ($t_{2(n-1)}$) and final ($t_{2n-1}$) intensities in an active phase. Thus,

$$\Delta I_n = I(t_{2n-1}) - I(t_{2(n-1)}) = I_{sat.} \frac{\left(1 - e^{-T/\tau_s}\right)\left(1 + e^{-(n-1)T/\tau_s}\right)}{1 + e^{-T/\tau_s}} \tag{S3}$$

For $n \to \infty$, the system reaches a dynamic steady state where the same response is obtained in each cycle. As a result, the signal amplitude in the steady state $\Delta I$ is given by

$$\Delta I_{n\to\infty} = I_{sat.} \frac{1 - e^{-T/\tau_s}}{1 + e^{-T/\tau_s}} = I_{sat.} \tanh\left(\frac{T}{2\tau_s}\right) \tag{S4}$$

## II. DERIVATION OF ENHANCEMENT FACTOR IN MSS

Suppose we have an MSS[2] signal $V(t)$ with normally distributed noise $\sigma$. The duration of each injection and purge phase is $T$, with $N$ repetitive measurements. A previous study showed that the signal response of the MSS can be modeled by a sorption-induced viscoelastic solution. Accordingly, $V(t)$ in the repetitive measurement is given by[1]

$$V(t) = \begin{cases} 0, & t = 0 \quad \text{(S5a)} \\ V_{sat.}\left[1 - \alpha \cdot e^{-\frac{t-t_0}{\tau_s}} \sum_{i=0}^{2(n-1)} \left(-e^{\frac{T}{\tau_s}}\right)^i - (1-\alpha)\cdot e^{-\frac{t-t_0}{\tau_r}} \sum_{i=0}^{2(n-1)} \left(-e^{\frac{T}{\tau_r}}\right)^i\right], & t_{2(n-1)} \le t < t_{2n-1} \quad \text{(S5b)} \\ V_{sat.}\left[\alpha \cdot e^{-\frac{t-t_0}{\tau_s}} \sum_{i=0}^{2n-1} \left(-e^{\frac{T}{\tau_s}}\right)^i + (1-\alpha)\cdot e^{-\frac{t-t_0}{\tau_r}} \sum_{i=0}^{2n-1} \left(-e^{\frac{T}{\tau_r}}\right)^i\right], & t_{2n-1} \le t < t_{2n} \quad \text{(S5c)} \end{cases}$$

with

$$\alpha = \frac{1}{\tau_s}\left(\frac{E_U}{E_R} - \frac{\tau_s}{\tau_r}\right)\left(\frac{1}{\tau_s} - \frac{1}{\tau_r}\right)^{-1}, \tag{S6}$$

$$V_{sat.} = \gamma E_R \lambda K_p C_g, \tag{S7}$$

where $\tau_s$ is the sorption time constant, $E_U$ is the unrelaxed modulus, $E_R$ is the relaxed modulus, $\tau_r$ is the stress relaxation time constant, $V_{sat.}$ is the saturation response, $\gamma$ is the stress-voltage transduction factor of the sensor, $\lambda$ is the sorption-induced swelling factor, $K_p$ is the Henry's constant, and $C_g$ is the gas concentration.

The signal amplitude of a single repetition as the signal approaches steady state ($t \to \infty$) is given by

$$\begin{aligned} \Delta V &= V(t_{2n-1}) - V(t_{2(n-1)}) \\ &= V_{sat.}\left[\alpha \tanh\left(\frac{T}{2\tau_s}\right) + (1-\alpha)\tanh\left(\frac{T}{2\tau_r}\right)\right] \end{aligned} \tag{S8}$$

The signal-to-noise ratio of a single measurement is defined as

$$\mathrm{SNR}_1 = \frac{\Delta V}{\sigma} \tag{S9}$$

As the number of repetitive measurements grows, according to the central limit theorem, the SNR is given by

$$\mathrm{SNR}_N = \frac{\Delta V}{\left(\frac{\sigma}{\sqrt{N}}\right)} = \frac{\Delta V \sqrt{N}}{\sigma}, \tag{S10}$$

The enhancement factor $E(N)$ due to the repetitive measurement is described as

$$E(N) = \frac{\mathrm{SNR}_N}{\mathrm{SNR}_1} = \sqrt{N}\left[\alpha \tanh\left(\frac{T}{2\tau_s}\right) + (1-\alpha)\tanh\left(\frac{T}{2\tau_r}\right)\right] \tag{S11}$$

which is independent of the noise level $\sigma$.

If the total measurement time is finite and fixed at $2\tau$, so that $2\tau = 2NT$, the enhancement factor is rewritten as

$$E(N) = \sqrt{N}\left[\alpha \tanh\left(\frac{\tau}{2\tau_s N}\right) + (1-\alpha)\tanh\left(\frac{\tau}{2\tau_r N}\right)\right] \tag{S12}$$

## III. ENHANCEMENT THRESHOLD AND THE TWO REGIMES

Repetitive averaging improves the SNR only when the enhancement factor exceeds unity for some $N$, i.e., when $E_{\max} \equiv \max_N E(N) > 1$. Since $E(N)$ depends on the repetition number only through the dimensionless ratios of the total measurement time to the intrinsic time constants, whether averaging is beneficial at all is governed by these ratios rather than by $N$.

For the single-relaxation model of the main text [Eq. (5)], $E(N) = \sqrt{N}\tanh(\tau/2\tau_s N)$ depends on the single ratio $\tau/\tau_s$. As $N$ increases from unity, $E(N)$ first rises (since tanh is approximately linear for small arguments, giving $E \sim \sqrt{N}\cdot(\tau/2\tau_s N) \propto 1/\sqrt{N}\cdot(\tau/\tau_s)$ in the strongly subdivided limit), and the trade-off between the $\sqrt{N}$ noise gain and the tanh signal loss produces a single interior maximum at an optimal $N^*$. The peak value $E_{\max}$ increases monotonically with $\tau/\tau_s$, and crosses unity at

$$(\tau/\tau_s)_{\text{th}} \approx 3. \tag{S13}$$

Below this threshold the response is too slow to develop within any subdivision of the available time, so $E(N) < 1$ for every $N$ and repetition cannot improve the SNR. Above it, averaging is beneficial within a finite window $1 \leq N \leq N^*$.

For the viscoelastic model of Eq. (S12), the signal build-up is governed by two time constants ($\tau_s$, $\tau_r$) and the unrelaxed-to-relaxed modulus ratio $E_U/E_R$ through the coupling factor $\alpha$. The enhancement threshold therefore shifts with the relaxation dynamics and the modulus ratio, and is generally lower than the single-relaxation value because the fast relaxation channel ($\tau_r \ll \tau_s$) contributes additional early-time signal. The threshold is obtained numerically as the value of $\tau/\tau_s$ at which $E_{\max} = 1$ for the fitted $\tau_s/\tau_r$ and $E_U/E_R$, and is mapped in Fig. S1.

Figure S1 shows the peak enhancement $E_{\max}$ over the $(\tau_s/\tau_r,\ \tau/\tau_s)$ plane at $E_U/E_R = 8$. The $E_{\max} = 1$ contour is the enhancement threshold. The threshold rises with $\tau_s/\tau_r$ and drops as $E_U/E_R$ increases (dashed guides for $E_U/E_R = 4$ and 16), confirming that it depends on both the relaxation dynamics and the modulus ratio rather than on $\tau/\tau_s$ alone.

Table S1 lists the parameters extracted from a single measurement for the two receptor–analyte systems used in the experiment, together with the resulting thresholds and enhancement metrics.

The two systems were deliberately chosen to bracket the enhancement transition. The fast-sorption case (poly($\varepsilon$-caprolactone) (PCL) exposed to $n$-nonane) lies well above its threshold and exhibits a pronounced maximum, whereas the slow-sorption case (poly(methyl methacrylate) (PMMA) exposed to methanol) sits essentially at its threshold, so that within the measurement noise $E(N)$ shows no meaningful enhancement. Their operating points are marked in Fig. S1, falling on opposite sides of the $E_{\max} = 1$ contour.

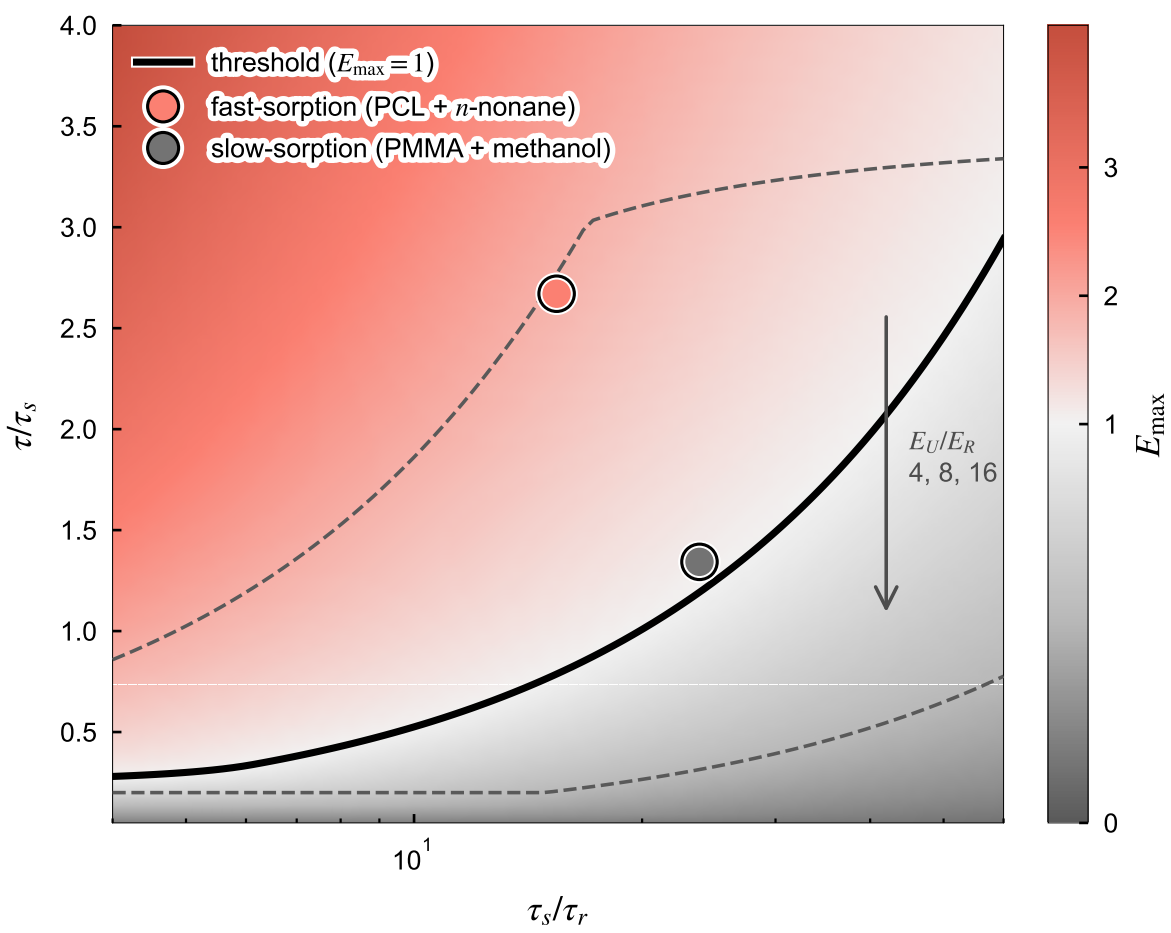

FIG. S1. Enhancement threshold of the viscoelastic model. The color map shows the peak enhancement factor $E_{\max} = \max_N E(N)$ over the $(\tau_s/\tau_r,\ \tau/\tau_s)$ plane at $E_U/E_R = 8$. The solid black curve is the $E_{\max} = 1$ threshold. Dashed guides show the threshold for $E_U/E_R = 4$ and 16 (the arrow indicates that increasing the modulus ratio lowers the threshold). Points mark the two experimental receptor–analyte systems (both with $E_U/E_R \approx 8$), which bracket the transition.

## IV. EXPERIMENTAL DETAILS

We constructed a custom measurement platform to evaluate our theoretical predictions. The system includes a polymer-coated MSS, two mass flow controllers (MFCs), a gas cylinder, and a bubbling vial for sample liquids, as shown in Fig. S2.

The analyte concentration was controlled by the two MFCs arranged as a dilution line. The carrier gas was split into two streams. One MFC regulated the flow directed through the bubbling vial, where the carrier stream was saturated with analyte vapor, while the other MFC regulated a bypass flow of pure carrier gas. The two streams were recombined before reaching the sensor cell, so that the analyte concentration is set by the ratio of the saturated flow to the total flow. Keeping the total flow rate constant, the concentration is therefore tuned solely through the flow-rate ratio of the two MFCs, which avoids concurrent changes in the total flow and the associated flow-induced baseline. In the present measurements the analyte concentration was set to 10% of the saturated vapor pressure by holding the saturated flow at one-tenth of the total flow (e.g., 10 sccm through the bubbling vial and 90 sccm of pure carrier gas at a total flow of 100 sccm). During each injection phase the analyte-laden mixture was delivered at this concentration, and during each purge phase pure carrier gas was supplied. The corresponding absolute concentrations listed in Table S1 were obtained from the saturated vapor pressure at 25 °C computed with Antoine coefficients from Ref.[3].

TABLE S1. Parameters extracted from a single measurement and the derived enhancement metrics for the two receptor–analyte systems ($2\tau$ = 360 s).

| | fast-sorption case | slow-sorption case |
|---|---|---|
| Receptor | PCL | PMMA |
| Analyte | *n*-nonane | methanol |
| Concentration [% $P_{\mathrm{sat}}$] | 10 | 10 |
| Concentration [ppmv] | $5.7 \times 10^2$ | $1.7 \times 10^4$ |
| $E_U/E_R$ | 8.5 | 7.9 |
| $\tau_s$ [s] | 67 | 134 |
| $\tau_r$ [s] | 4.4 | 5.6 |
| $\tau/\tau_s$ | 2.7 | 1.3 |
| threshold $(\tau/\tau_s)_{\mathrm{th}}$ | 0.7 | 1.2 |
| $E_{\mathrm{max}}$ | 1.9 | 1.1 |
| $N^*$ | 18 | — |

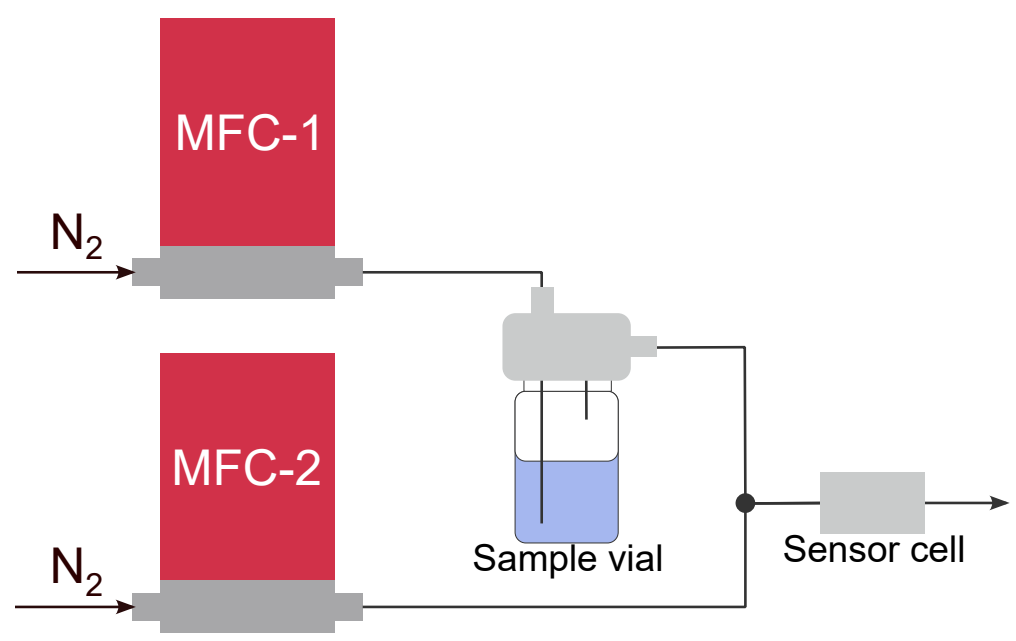

FIG. S2. Schematic of the measurement setup.

Each experiment was conducted under an identical total measurement time $2\tau = 360$ s, varying only the repetition number $N$ from 1 to 90. Two receptor–analyte systems were measured to probe the two regimes discussed above. For the fast-sorption case, a PCL-coated MSS was exposed to $n$-nonane vapor. For the slow-sorption case, a PMMA-coated MSS was exposed to methanol vapor. Here a small number of cycles were corrupted by transient acquisition faults and were re-measured under identical experimental conditions (same chip, receptor, analyte, and flow settings) on a separate day, and the re-measured cycles were substituted for the faulty ones before analysis.

## V. DATA PROCESSING

To quantify the signal amplitude and the noise level, we first folded the record onto a single cycle and averaged over the $N$ repetitions,

$$\bar{V}_N(t) = \frac{1}{N}\sum_{n=1}^{N} V(t + 2(n-1)T), \quad 0 \le t < 2T, \tag{S14}$$

where each cycle comprises an active (injection) stage $0 \le t < T$ followed by a purge stage $T \le t < 2T$, and is offset so that $\bar{V}_N(0) = 0$.

The noise level is evaluated on a short window near the end of the purge stage, where the averaged response has settled. Over a window $W$ of width $\delta = 0.2$ s centered 1 s before the end of the cycle, we remove a linear trend $\tilde{V}_N(t)$ (least-squares fit over $W$) and take the RMS of the residual,

$$\bar{n}(t) = \bar{V}_N(t) - \tilde{V}_N(t), \quad t \in W, \tag{S15}$$

$$\sigma_N = \sqrt{\frac{1}{M}\sum_{t_i \in W} \bar{n}(t_i)^2}, \tag{S16}$$

where $M$ is the number of samples in $W$. Detrending removes any residual baseline drift so that $\sigma_N$ reflects the random noise of the averaged trace.

The signal amplitude is taken as the saturated plateau reached at the end of the active stage, evaluated as the mean of $\bar{V}_N$ over the last $\delta = 0.2$ s of the injection window,

$$\Delta V_N = \frac{1}{M}\sum_{T-\delta \le t_i < T} \bar{V}_N(t_i). \tag{S17}$$

As a result, the SNR is given by

$$\mathrm{SNR}_N = \frac{\Delta V_N}{\sigma_N} \tag{S18}$$

The enhancement factor for the experimental data is calculated by

$$E(N) = \frac{\mathrm{SNR}_N}{\mathrm{SNR}_1} \tag{S19}$$

The experimental results as well as the data processing procedure are illustrated in Fig. S3.

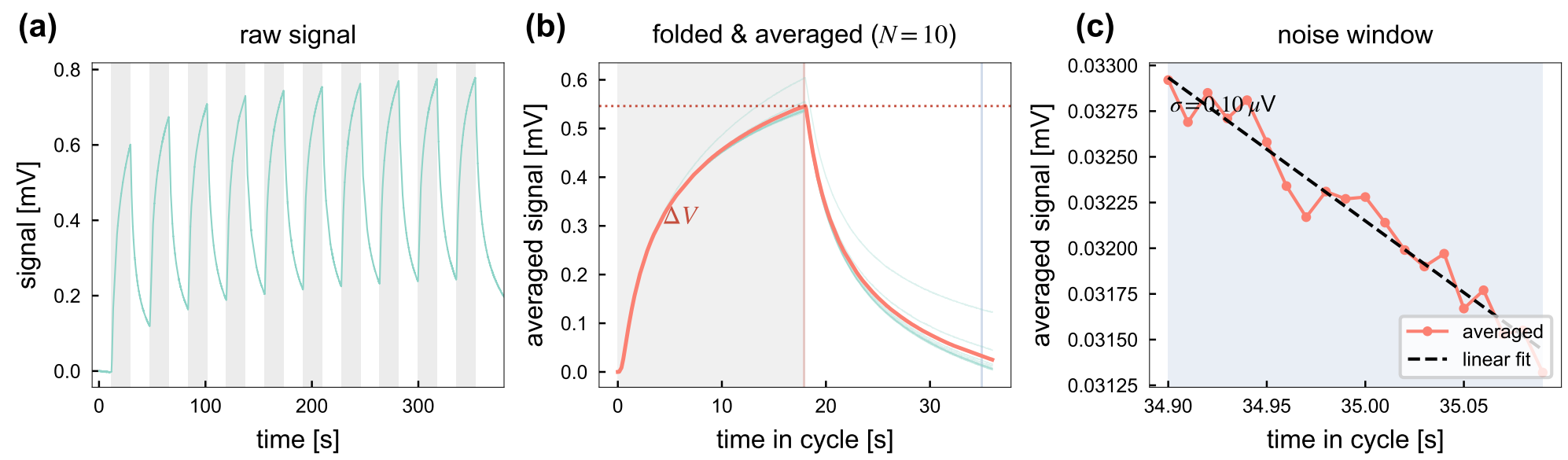

FIG. S3. Data processing procedure, illustrated for the fast-sorption case (PCL, $N = 10$). (a) Raw sensor response over the full record. The injection half of each cycle is shaded. (b) Response folded onto a single cycle and averaged over the $N$ repetitions. The signal amplitude $\Delta V_N$ is the plateau at the end of the injection stage (red band), and the noise window near the end of the purge stage is marked (blue band). (c) Zoom on the noise window. The residual of the averaged trace about a linear fit (dashed) gives $\sigma_N$.

## REFERENCES

[1] K. Minami, K. Shiba, and G. Yoshikawa, "Sorption-induced static mode nanomechanical sensing with viscoelastic receptor layers for multistep injection-purge cycles," J. Appl. Phys. **129**, 124503 (2021).

[2] G. Yoshikawa, T. Akiyama, S. Gautsch, P. Vettiger, and H. Rohrer, "Nanomechanical Membrane-type Surface Stress Sensor," Nano Lett. **11**, 1044–1048 (2011).

[3] C. L. Yaws, *The Yaws Handbook of Vapor Pressure: Antoine Coefficients*, second edition ed. (Gulf Professional Publishing is an imprint of Elsevier, Kidlington, Oxford Waltham, MA, 2015).